\documentclass[%
 aip,
 amsmath,amssymb,
 reprint,%
]{revtex4-1}

\usepackage{graphicx}
\usepackage{dcolumn}
\usepackage{bm}

\usepackage[utf8]{inputenc}
\usepackage[T1]{fontenc}

\usepackage{xcolor}
\usepackage{cancel}
\usepackage[uline]{hhtensor}
\usepackage{tikz}
\usepackage{mathrsfs}

\usepackage{float}

\definecolor{myblue}{rgb}{0., 0.4, 1.0} 
\definecolor{myred}{rgb}{0.831, 0.0, 0.0}

\usepackage{draftwatermark}
\SetWatermarkText{\bf PREPRINT -- PREPRINT -- PREPRINT}
\SetWatermarkFontSize{4cm}
\SetWatermarkAngle{90}
\SetWatermarkHorCenter{20.5cm}
\SetWatermarkColor[gray]{0.7}

\newcommand{\redline}{\raisebox{2pt}{\tikz{\draw[-,black!0!red,solid,line width = 1.2pt](0,0) -- (5.5mm,0);}}}

\newcommand{\myblueline}{\raisebox{2pt}{\tikz{\draw[-,black!0!myblue,solid,line width = 1.2pt](0,0) -- (5.3mm,0);}}}

\newcommand{\bluedashedline}{\raisebox{2pt}{\tikz{\draw[-,black!0!blue,dashed,line width = 1.2pt](0,0) -- (5.3mm,0);}}}

\newcommand{\myreddashedline}{\raisebox{2pt}{\tikz{\draw[-,black!0!myred,dashed,line width = 1.2pt](0,0) -- (5.3mm,0);}}}

\begin{document}

\preprint{AIP/123-QED}

\title{A 14-moment maximum-entropy description of electrons in crossed electric and magnetic fields}





\author{S.~Boccelli}
  \email{Author to whom correspondence should be addressed: stefano.boccelli@polimi.it}
  \affiliation{Department of Aerospace Science and Technology, Politecnico di Milano, via La Masa 34 I-20156 Milano, Italy.}
 
\author{F.~Giroux}%
  \email{fgiro059@uottawa.ca}
  \affiliation{Department of Mechanical Engineering, University of Ottawa, Ottawa, Ontario K1N 6N5, Canada.}

\author{T.~E.~Magin}
  \email{thierry.magin@vki.ac.be}
  \affiliation{Aeronautics and Aerospace department, von Karman Institute for Fluid Dynamics, Waterloosesteenweg 72 B-1640 Sint-Genesius-Rode, Belgium.}

\author{C.~P.~T.~Groth}
  \email{groth@utias.utoronto.ca}
  \affiliation{University of Toronto Institute for Aerospace Studies, 4925 Dufferin Street, Toronto, Ontario M3H 5T6, Canada.}

\author{J.~G.~McDonald}
  \email{james.mcdonald@uottawa.ca}
  \affiliation{Department of Mechanical Engineering, University of Ottawa, Ottawa, Ontario K1N 6N5, Canada.}


\date{\bf Accepted in Physics of Plasmas, {\bf 27}, 2020.}

\begin{abstract}
ABSTRACT\\
A 14-moment maximum-entropy system of equations is applied to the description of non-equilibrium electrons
in crossed electric and magnetic fields and in the presence of low collisionality, characteristic of low-temperature plasma devices.  The flexibility of this formulation is analyzed through comparison with analytical
results for steady-state non-equilibrium velocity distribution functions and against particle-based
solutions of the time-dependent kinetic equation.  Electric and magnetic source terms are derived for the
14-moment equations, starting from kinetic theory.  A simplified BGK-like collision term is formulated to describe the collision of electrons with background neutrals, accounting for the large mass disparity and for energy exchange.
An approximated expression is proposed for the collision frequency, to include the effect of the electrons drift velocity, showing good accuracy in the considered conditions.
The capabilities of the proposed 14-moment closure to capture accurately non-equilibrium behaviour of electrons for space homogeneous problems under conditions representative of those found in Hall thrusters is demonstrated.
\end{abstract}

\maketitle


\section{Introduction}

Electrons in crossed electric and magnetic fields show strongly non-equilibrium distribution functions when the collisionality is low (high Hall parameter) and the $\bm{\mathsf{E}}\times \bm{\mathsf{B}}$ drift velocity is comparable or larger than the electrons thermal speed.\cite{taccogna2016non}
This is the case, for example, in Hall Effect Thruster (HET) devices used for space propulsion (also known as Stationary Plasma Thrusters - SPT),\cite{zhurin1999physics,morozov2000fundamentals,boeuf2017tutorial} and in magnetron devices.\cite{shon2002modeling,sheridan1998electron}

\begin{figure}[htpb]
  \centering
  \includegraphics[width=0.55\columnwidth]{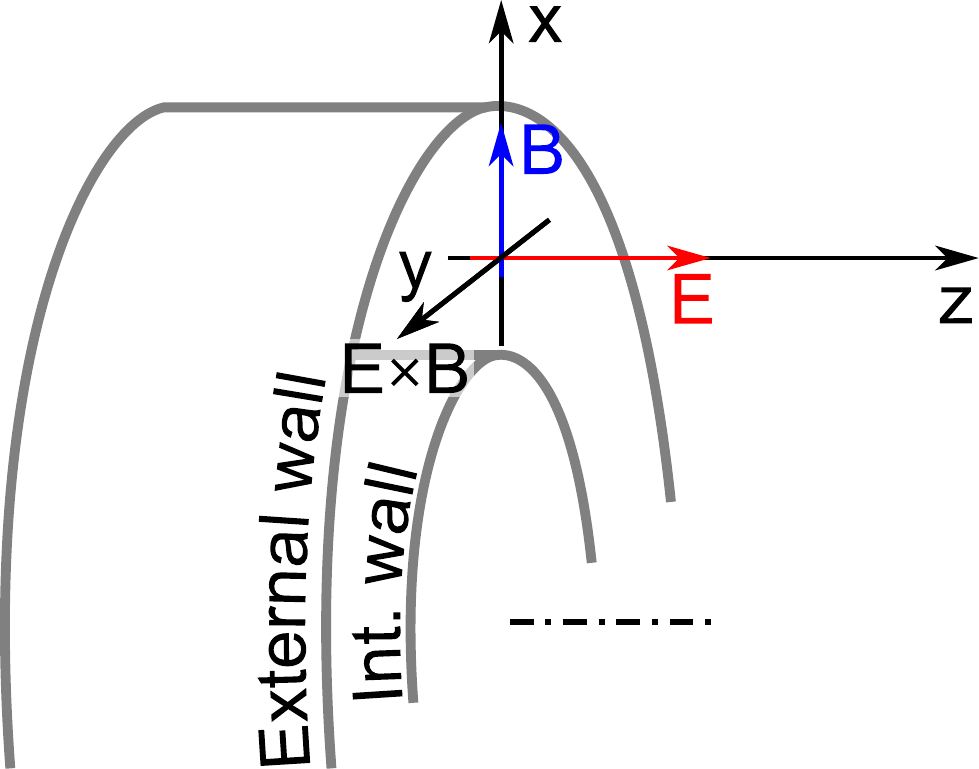}
  \caption{Schematic view of crossed electric field $\mathsf{E}$ and magnetic field $\mathsf{B}$ at the exit plane of a Hall thruster geometry, and reference system employed in this work.}
  \label{fig:ref-system-thruster}
\end{figure}

In the simplest configuration, the magnetic field in a Hall thruster is predominantly radial, and the electric field, which is responsible for the acceleration of ions, is mostly longitudinal, as sketched in Fig.~\ref{fig:ref-system-thruster}, although other components can appear due to three-dimensional effects and plasma instabilities.\cite{boeuf2017tutorial}

The topic of thermodynamic non-equilibrium transport of electrons in Hall thrusters has been subjected to broad investigation in the electric propulsion community, from the theoretical, experimental and numerical perspectives.\cite{fedotov1999electron,barral2000double,shagayda2012stationary,guerrini1997characterization,lago1997electron,shimura1993electron}
In the operating conditions of interest for Hall thrusters, Coulomb collisions are often negligible compared to collisions with the neutral background.\cite{morozov2000fundamentals}
For high values of the Hall parameter, when the electron cyclotron frequency is much higher than the frequency of electron collisions with background neutrals, one can show that the trochoid motion of electrons in the crossed $\bm{\mathsf{E}}$ and $\bm{\mathsf{B}}$ fields gives rise to a ring-shaped velocity distribution function (VDF) centered around the drift velocity, $u_d = \mathsf{E}/\mathsf{B}$.\cite{shagayda2012stationary}
In terms of the energy distribution function (EDF), two peaks become clearly visible, one associated with the portion of the trajectory with minimum velocity and one with the fast portion of the trochoid.\cite{fedotov1999electron,barral2000double}
Considering the presence of electron-neutral collisions, ionizing reactions also affect the shape of the distribution function by attenuating the high-energy region and producing a population of colder secondary electrons.\cite{shagayda2012stationary,vahedi1995monte}
In contrast, elastic collisions with background neutrals randomize the velocities, and thus tend to make the distribution isotropic.
The steady-state shape for the distribution function will arise as a balance between these effects, whose relative importance can be described by the Hall parameter.

In the context of electric propulsion, non-equilibrium in the electrons EDF is also affected by the presence of solid boundaries.
As walls charge negatively and a plasma sheath is created, only electrons above a certain energy are absorbed and low-energy electrons are reflected back into the plasma.
Moreover, cold secondary electrons are emitted by ceramic walls, and their effect on the electrons distribution function and conductivity has been investigated by a number of authors.\cite{kaganovich2007kinetic,morozov2001theory}
In this work, we neglect any wall interaction and only consider the effects of collisionality and crossed $\bm{\mathsf{E}}$ and $\bm{\mathsf{B}}$ fields.

The non-equilibrium state of an electron population can be properly described in the framework of kinetic theory of gases and plasmas.
Solving the Vlasov equation (supplemented with a suitable collision operator), one can obtain the time evolution of the VDF in phase space.\cite{montgomery1964plasma}
A numerical solution to this problem is often tackled with the Particle-In-Cell (PIC) method.\cite{birdsall2018plasma}
However, the rapid timescales introduced by the plasma frequency, together with the extremely small spatial scales required to resolve the Debye length often make this method computationally demanding.
In particular, for the operating conditions of Hall thrusters, even simplified two-dimensional simulations often require weeks or months of computational time on parallel architectures. \cite{charoy20192d} 
Obtaining fully consistent three-dimensional results proves to be a formidable task and is currently possible only by employing scalings that only guarantee a partial similarity.\cite{taccogna2011three,szabo2014full}

As opposed to kinetic theory, fluid-based approaches such as the multi-fluid description are characterized by very affordable computational times. 
However, classical fluid dynamic descriptions, such as the Euler and Navier-Stokes-Fourier systems of equations, fail to provide accurate predictions in strong non-equilibrium situations, as transport quantities appearing in the fluxes are not correctly reproduced for highly non-Maxwellian VDFs.
In other words, neglecting the shear stresses and heat flux (Euler equations) or approximating them by the continuum assumption and the Fourier's law for the heat flux (Navier-Stokes-Fourier equations) constitute questionable assumptions and prove to be theoretically justified and accurate only for near-equilibrium situations.\cite{ferziger1972mathematical,lofthouse2008nonequilibrium}

In the context of gas dynamics and atmospheric entry, a number of strategies have been developed for extending the fluid description towards non-equilibrium conditions,
for example in terms of moment methods\cite{struchtrup2005macroscopic} or extended thermodynamics.\cite{muller1993extended}
A large portion of such descriptions is perturbative:
the distribution function is written as a Maxwellian at the local conditions of density, velocity and temperature, perturbed by a series of polynomials of a given order.\cite{ferziger1972mathematical}
The Chapman-Enksog expansion and the Grad method\cite{grad1949kinetic} are two examples of such perturbative approach and naturally reduce to the Navier-Stokes-Fourier and the Euler equations when perturbations are respectively small or negligible.
These schemes have been also formalized in the context of plasma physics by a number of authors.\cite{braginskii1965transport,zhdanov2002transport,graille2009kinetic} 
However, the range of non-equilibrium covered by these methods is relatively limited, due to the choice of using a Maxwellian as the starting point for the perturbative approach.
Indeed, in order to reproduce strong non-equilibrium conditions, the required order of the expansion often becomes quite large.\cite{khazanov2010kinetic,torrilhon2015convergence} This brings additional complexity and numerical cost.
Moreover, artificial regularization strategies may be needed to rule out unphysical behaviors,\cite{struchtrup2003regularization} but at the price of introducing arbitrary assumptions and losing the hyperbolic structure of the system.
Finally, we should mention another strategy widely used in plasma chemistry, namely the two-term approximation of the Boltzmann equation,\cite{capitelli2015fundamental} based on an expansion around a local symmetric distribution, together with its multi-term extensions.\cite{loffhagen1996two}

\begin{figure}[htpb]
    \centering
    \includegraphics[width=0.9\columnwidth]{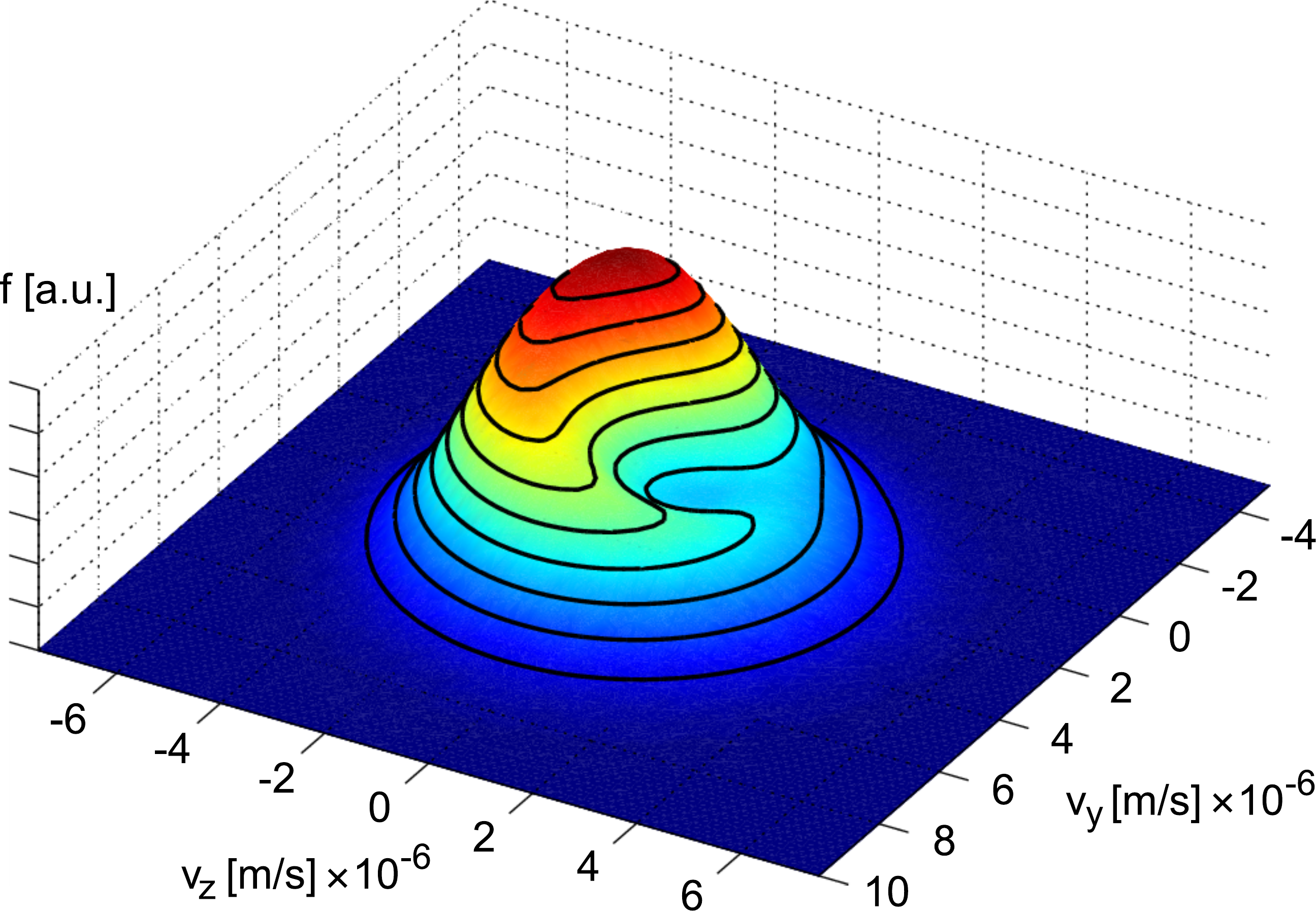}
    \caption{Example of an electron VDF in crossed electric and magnetic fields, obtained with a simplified BGK collision operator, from the model of Shagayda\cite{shagayda2012stationary} with parameters $T_b = 10^5$ K, $\beta = 0.3$, $\mathsf{E/B} = 2.5 \times 10^6$ m/s. 
    Velocity components $v_y$ and $v_z$ are perpendicular to the magnetic field.}
    \label{fig:VDF-electrons-example}
\end{figure}

A moment description for low-collisional and magnetized plasmas should, first of all, include temperature and pressure anisotropy, arising from the decoupling between the direction parallel and perpendicular to the magnetic field.\cite{zhdanov2002transport}
In the case of electrons in crossed $\bm{\mathsf{E}}$ and $\bm{\mathsf{B}}$ fields one can observe further features.
To fix the ideas, we show in Fig.~\ref{fig:VDF-electrons-example} as an example a steady-state VDF for low-collisional electrons, obtained by solving the simplified kinetic equation employed by Shagayda\cite{shagayda2012stationary} (see test case in Section \ref{sec:0D-relaxation}).
Such model assumes that electron-neutral collisions bring the electron population towards a Maxwellian distribution with temperature $T_b$, and the collision rate is obtained from the inverse Hall parameter $\beta$. 
We refer the reader to the original paper for further details.
Figure~\ref{fig:VDF-electrons-example} shows the resulting VDF in the perpendicular plane $(v_y, v_z)$, and clearly shows additional anisotropy and asymmetry.
As a result, most of the odd-order moments (off-diagonal components of the pressure tensor and the heat flux) will, in general, be non-zero.
Moreover, when the electric drift velocity happens to be comparable or larger than the electrons thermal velocity, as in the case of Fig.~\ref{fig:VDF-electrons-example}, the distribution will show bi-modal (``ring-like'' in 2D) shapes, making higher even-order moments more pronounced by shifting the density from the average value towards the tails.
Clearly, the relative importance of the phenomena listed above varies according to the collisionality, fields strength, and electrons temperature.

In attempting to develop a moment description for a non-equilibrium situation, achieving an exact reproduction of the distribution function itself is not strictly necessary, as long as its moments can be
reproduced with sufficient accuracy.  However, the description should be able to reproduce the main features mentioned above.

\subsection{Proposed description and structure of this work}

In this work, we investigate a non-perturbative strategy, based on the maximum-entropy family of moment methods.\cite{levermore1996moment,muller1993extended}
Such formulations allow one to naturally describe large deviations from equilibrium, assuming a shape for the distribution function that maximises the entropy for a given set of moments.
This approach provides a distribution function which is naturally bounded and always positive, avoiding a number of negative characteristics of perturbative descriptions. 
Moreover, the resulting system of moments can be proven to be hyperbolic whenever the underlying entropy-maximization problem is solvable.\cite{levermore1996moment,junk1998}

While the various mathematical and computational benefits of maximum-entropy closures have been demonstrated in several recent studies \cite{groth:2009,mcdonald:2014}, the high-order members of this closure hierachy have long been perceived as being exceedingly expensive for practical applications from the overall computational standpoint.  This is due to the need to perform sub-iterations in order to find the maximum-entropy distribution, from which transport fluxes can then be computed.
This would nullify all computational gain of this fluid model, with respect to a full kinetic description of the problem.
However, recent developments have provided approximated interpolative approaches to the entropy maximisation problem, which allow for an agile and affordable solution to be found for selected systems of equations,\cite{mcdonald2013affordable} and their application to multi-dimensional non-equilibrium gaseous flows has yielded very promising results.\cite{tensuda:2014,tensuda:2016}

In this work, we consider a 14-moment maximum-entropy description for the electron population.\cite{kremer1986extended,muller1993extended,mcdonald2013affordable}
First, in Section~\ref{sec:kinetic-framework-max-ent} we discuss the kinetic equation and provide a brief description of the maximum-entropy framework, together with the system of 14 moments adopted here.
The flexibility of the 14-moment system is then investigated in Section~\ref{sec:analytical-vs-14mom} by checking how well the maximum-entropy approximation can reproduce a set of analytical distribution functions for magnetized electrons.
As the present approximation proves accurate enough, we proceed to build the source terms required by the set of 14 moment equations.
In Section~\ref{sec:EM-fields-sources} the electric and magnetic terms are developed from the kinetic equation, and 
in Section~\ref{sec:collision-term}, a set of possible collision terms is developed for collisions of electron with background neutrals.
Finally, in Section~\ref{sec:0D-relaxation} the solution of the system of moment equations is compared to a kinetic particle-based simulation for the situation of the relaxation of an initially Maxwellian distribution towards its non-equilibrium steady state.  The current study is restricted to representative space-homogeneous or zero-dimensional (0D) problems with the aim of providing a baseline foundation of results upon which further multi-dimensional simulations can eventually be built.

This work targets low-temperature plasmas and only electron-neutral collisions are accounted for in this work. 
Higher collisionality than considered would increase the reliability of the scheme, as the system would remain closer to equilibrium.
Particular emphasis is put to operating conditions somewhat representative of Hall thruster discharges. 

It should be mentioned that different moment formulations have been already proposed for this particular case, both in the homogeneous case and in presence of space gradients, based on analytical results for the electrons distribution functions.\cite{shagayda2015electron,shagayda2017analytic}
Rather than finding a particular solution for the current problem, in this paper we aim at investigating the suitability of the 14-moment closure for this type of problem, in view of its general character and its possibilities for direct generalization.


\section{Moment description of electrons}\label{sec:kinetic-framework-max-ent}

\noindent We describe the electron population by the Vlasov kinetic equation supplemented by a collision term,\cite{montgomery1964plasma}
\begin{equation}\label{eq:kinetic-eq-vlasov}
  \frac{\partial f}{\partial t}
+ \bm{v} \cdot \frac{\partial f}{\partial \bm{x}} 
+ \frac{q}{m} \left(\bm{\mathsf{E}} + \bm{v} \times \bm{\mathsf{B}} \right) \cdot \frac{\partial f}{\partial \bm{v}}
= \mathscr{C} \, ,
\end{equation}

\noindent with $q$ and $m$ the electron charge and mass, respectively, $\mathsf{E}$ and $\mathsf{B}$ the electric and magnetic fields, $\bm{x}$ and $\bm{v}$ the space and velocity coordinates, $f(v)$ the electrons velocity distribution function and $\mathscr{C}$ the collision operator, to be defined in the following.

A fluid description is obtained by multiplying the kinetic equation by proper particle quantities such as mass, momentum and energy, and integrating over the velocities.
Defining a function of the velocity $\phi(v)$, one obtains the general moment equation\cite{ferziger1972mathematical}
\begin{multline}\label{eq:generalized-moment-eq}
  \frac{\partial \left( n \left< \phi \right> \right)}{\partial t}
+ \frac{\partial}{\partial \bm{x}} \cdot \left[ n \left< \bm{v}\phi \right> \right]
= 
\\
\frac{q}{m} \left< \left( \bm{\mathsf{E}} + \bm{v}\times \bm{\mathsf{B}} \right) \cdot \frac{\partial \phi}{\partial \bm{v}} \right>
+ \left< \phi \mathscr{C} \right> \, ,
\end{multline}

\noindent where $n$ is the electrons number density and the angle bracket operator represents the integration over velocity space,
\begin{equation}
    \left<  \bullet \right> \equiv \int\!\!\int\!\!\int_{-\infty}^{+\infty} \bullet \ f(v) \ \mathrm{d}^3 v \, .
\end{equation}

As can be appreciated from Eq.~(\ref{eq:generalized-moment-eq}), in every moment equation originating from a function $\phi(v)$ of order $p$ in the velocity, the flux term, $\left< \bm{v} \phi \right>$, will introduce moments of order $p+1$.
Therefore, an infinite hierarchy of such moment equations would be needed in order to solve the problem.
Practically speaking, one truncates this hierarchy to a desired number of moment equations and postulates a closure for all the higher-order ``closing'' moments.


\subsection{Maximum-entropy descriptions}

In the maximum-entropy framework, the closure for the additional moments is found by assuming that the distribution function is the one that maximises the entropy for a set of known moments.
Such a distribution function takes the form of the exponential of a polynomial,\cite{levermore1996moment} 
\begin{equation}\label{eq:max-ent-VDFs}
    f(v) = \exp \left[ \bm{\alpha}^\top \bm{\Phi}(v) \right] \,,
\end{equation}

\noindent where $\bm{\alpha}$ is a vector of weights and $\bm{\Phi}$ is a vector of monomial functions of the particles velocity, $v$. 
In this work, we consider a maximum degree of 4.
To fix the ideas, considering one only degree of freedom for particle motion (``1D physics'') the maximum-entropy distribution would read
\begin{equation}
    f(v) = \exp\left(\alpha_0 + \alpha_1 v + \alpha_2 v^2 + \alpha_3 v^3 + \alpha_4 v^4 \right) \, .
\end{equation}

\noindent Such a shape reduces to a centered Maxwellian when only the coefficients $\alpha_0$ and $\alpha_2$ are non-zero, but allows for a number of strongly non-equilibrium distributions to arise, such as the Druyvesteyn distribution\cite{lieberman2005principles,druyvesteyn1940mechanism} when only $\alpha_0$ and $\alpha_4$ are non-zero, plus all intermediate situations.
Considering a full 3D case, the minimum set of generating functions that allows for a well behaved fourth-order maximum-entropy distribution is\cite{levermore1996moment}
\begin{equation}\label{eq:phi-gen-functions}
    \bm{\Phi} = m \left(1, v_i, v_i v_j, v_i v^2, v^4 \right) \, .
\end{equation}

\noindent This vector is composed of 14 terms, considering the three components of the particles velocity $v_i$, with $i \in \{x,y,z\}$.
If compared to a Maxwellian distribution, this VDF includes the possibility of anisotropy in the temperatures through the $v_i v_j$ entries, it can be asymmetric and thus have an heat flux due to the $v_i v^2$ third order term and has the possibility of presenting bi-modal shapes, thanks to the $v^4$ term.
These features make this VDF interesting for the considered non-equilibrium distributions of Fig.~\ref{fig:VDF-electrons-example}.

For every entry in the vector $\bm{\Phi}$, Eq.~(\ref{eq:generalized-moment-eq}) allows one to obtain a moment equation, resulting in a system of 14 balance laws describing the (non-equilibrium) electron fluid,
\begin{equation}\label{eq:governing-equations-fluid}
    \frac{\partial \bm{U}}{\partial t} + 
    \bm{\nabla} \cdot \bm{F} =
    \bm{S}_{em} + \bm{S}_c \, ,
\end{equation}

\noindent where $\bm{U}$ is the state vector, $\bm{F} = \bm{F}(\bm{U}) = [\bm{F}_x, \bm{F}_y, \bm{F}_z]$ are the fluxes in the $x$, $y$, and $z$ directions while $\bm{S}_{em}$ and $\bm{S}_c$ are the electro-magnetic and collisional source terms. 
The full expression for the vector of conserved variables and the fluxes is reported in Appendix \ref{sec:appendix-full-sys14mom} and are only sketched here as

\begin{equation}\label{eq:Ui-definitions}
  \bm{U} 
= 
  \begin{pmatrix}
    U_1 \\
    U_2 \\
    U_3 \\
    U_4 \\
    U_5 \\
    U_6 \\
    U_7 \\
    U_8 \\
    U_9 \\
    U_{10} \\
    U_{11} \\
    U_{12} \\
    U_{13} \\
    U_{14} \\
  \end{pmatrix}
=
  \begin{pmatrix}
    \left< m \right> \\
    \left< m v_x \right> \\
    \left< m v_y \right> \\
    \left< m v_z \right> \\
    \left< m v_x v_x \right> \\
    \left< m v_x v_y \right> \\
    \left< m v_x v_z \right> \\
    \left< m v_y v_y \right> \\
    \left< m v_y v_z \right> \\
    \left< m v_z v_z \right> \\
    \left< m v_x v^2 \right> \\
    \left< m v_y v^2 \right> \\
    \left< m v_z v^2 \right> \\
    \left< m v^4 \right> \\
  \end{pmatrix}
=
  \begin{pmatrix}
    \rho \\
    \rho u_x \\
    \rho u_y \\
    \rho u_z \\
    \rho u_x u_x + P_{xx} \\
    \rho u_x u_y + P_{xy} \\
    \rho u_x u_z + P_{xz} \\
    \rho u_y u_y + P_{yy} \\
    \rho u_y u_z + P_{yz} \\
    \rho u_z u_z + P_{zz} \\
    \rho u_x u^2 + \cdots + q_x \\
    \rho u_y u^2 + \cdots + q_y \\
    \rho u_z u^2 + \cdots + q_z \\
    \rho u^4 + \cdots + R_{iijj} \\
  \end{pmatrix}
  \, .
\end{equation}

\noindent In such definition, the pressure tensor components $P_{ij}$, the heat flux vector components $q_i$ and the order-4 moment $R_{iijj}$ can be shown to be central moments, respectively of order two, three and four,
\begin{equation}
    P_{ij} = \left< m c_i c_j \right> \ \ , \  q_{i} = \left< m c_i c^2  \right> \ \ , \
    R_{iijj} = \left< m c^4 \right>\, ,
\end{equation}

\noindent with $c_i = v_i - u_i$ the peculiar velocity and $i \in \left\{x,y,z\right\}$.
Note that this definition of $q_i$ differs from standard fluid dynamics, where a factor $1/2$ is usually included.\cite{ferziger1972mathematical}
Source terms accounting for electro-magnetic fields are developed in Section~\ref{sec:EM-fields-sources}, and electron-neutral collision sources are described in Sections \ref{sec:BGK-electron-neutral} and \ref{sec:isotropic-E-preserving-collis}.

As anticipated, the vector of fluxes $\bm{F}$ contains some additional moments which are not described in the vector $\bm{U}$, and thus require a closure of some kind.\cite{mcdonald2013affordable}
The closing moments required are the 10 components of the symmetric heat flux tensor, $Q_{ijk}$, 6 components for the fourth-order moment, $R_{ijkk}$ and 3 components for the contracted fifth-order moment, $S_{ijjkk}$, defined as
\begin{subequations}
\begin{align}
    Q_{ijk} &= \left< m c_i c_j c_k \right> = \int m c_i c_j c_k f({v}) \, \mathrm{d}^3 v \, ,\\
    R_{ijkk} &= \left< m c_i c_j c^2 \right> = \int m c_i c_j c^2 f({v}) \, \mathrm{d}^3 v \, ,\\
    S_{ijjkk} &= \left< m c_i c^4 \right> = \int m c_i c^4 f({v}) \, \mathrm{d}^3 v \, ,
\end{align}
\end{subequations}

The vector $q_i$, appearing in Eq.~(\ref{eq:Ui-definitions}), is the contraction of $Q_{ijk}$.
In the traditional maximum-entropy framework, the closing moments are actually obtained by integrating the VDF over the velocity space during CFD computations.
This results in additional computational overhead.
Even most importantly, finding the distribution function $f(v)$ of Eq.~(\ref{eq:max-ent-VDFs}) for a given set of lower-order moments also proves to be a formidable task.
This is because no algebraic relation is known between the moments $\bm{U}$ and the vector of coefficients $\bm{\alpha}$ for moment closures of order 4 or higher.
For the case considered here, the relation between the two is indeed the integral of a fourth order polynomial in $\bm{\alpha}$, which one would need to invert.
The traditional approach employs iterative methods, and the problem can be stated as follows:\cite{levermore1996moment}
Given the target moments, $\bm{U}$, and a set of generating functions, $\bm{\Phi}$, from Eq.~(\ref{eq:phi-gen-functions}), we are interested in finding the weights $\bm{\alpha}$ such that
\begin{equation}\label{eq:minimization-problem}
    \bm{U} - \int \bm{\Phi} \exp\left[\bm{\alpha}^{\top} \bm{\Phi} \right] \, \mathrm{d}^3 v = \bm{0} \,.
\end{equation}

In this work, we solve the entropy-minimisation problem by a Newton search algorithm, where the function to be minimised is
\begin{equation}\label{eq:minimization-problem-J}
  J = \int \exp(\bm{\alpha}^{\top} \bm{\Phi}) \, \mathrm{d}^3 v - \bm{\alpha}^{\top} \bm{U} \, .
\end{equation}

Notice that the condition of zero gradient for $J$ retrieves Eq.~(\ref{eq:minimization-problem}).
By solving the minimisation problem, we are able to check whether the 14-moment maximum-entropy system can recover the desired kinetic solution in terms of distribution function and closing moments.
Therefore, this will be the preferred approach despite the additional overhead that this introduces in the computations. 

It is important to mention that approximated interpolative solutions to the entropy maximisation problem have also been developed for the order 4 maximum-entropy system.\cite{mcdonald2013affordable}
Such approach directly returns an approximated value of the closing moments in terms of the lower-order ones.
The approach bypasses completely both the costly iterations and the integration over the velocity space and allows one to obtain a very affordable solution.
This is in sharp contrast with the traditionally high computational cost of maximum-entropy systems.

Since this approach does not explicitly give the maximum-entropy VDF, we do not detail its application in the current work.
However, for all the computations performed here, the closing moments were checked with both the iterative procedure and the approximated closure, and were found to always provide very similar results.


\section{Comparison with analytical VDFs}\label{sec:analytical-vs-14mom}

As a first step towards assessing the flexibility of the 14-moment closure and its accuracy in the description of magnetized electrons, we consider the analytical VDFs obtained by Shagayda\cite{shagayda2012stationary} under the assumptions of steady state and uniform conditions.
Such results extend the analytical expressions of Fedotov et al.\cite{fedotov1999electron} and Barral et al.\cite{barral2000double} to the three velocity components of the distribution function and additionally introduce collisions in a BGK-like fashion.
From these VDFs, we compute, by numerical integration, the 14 moments $U_{1,\cdots,14}$, and the additional closing moments $Q_{ijk}$, $R_{ijkk}$ and $S_{ijjkk}$.

\begin{figure*}[htpb]
    \centering
    \includegraphics[width=\textwidth]{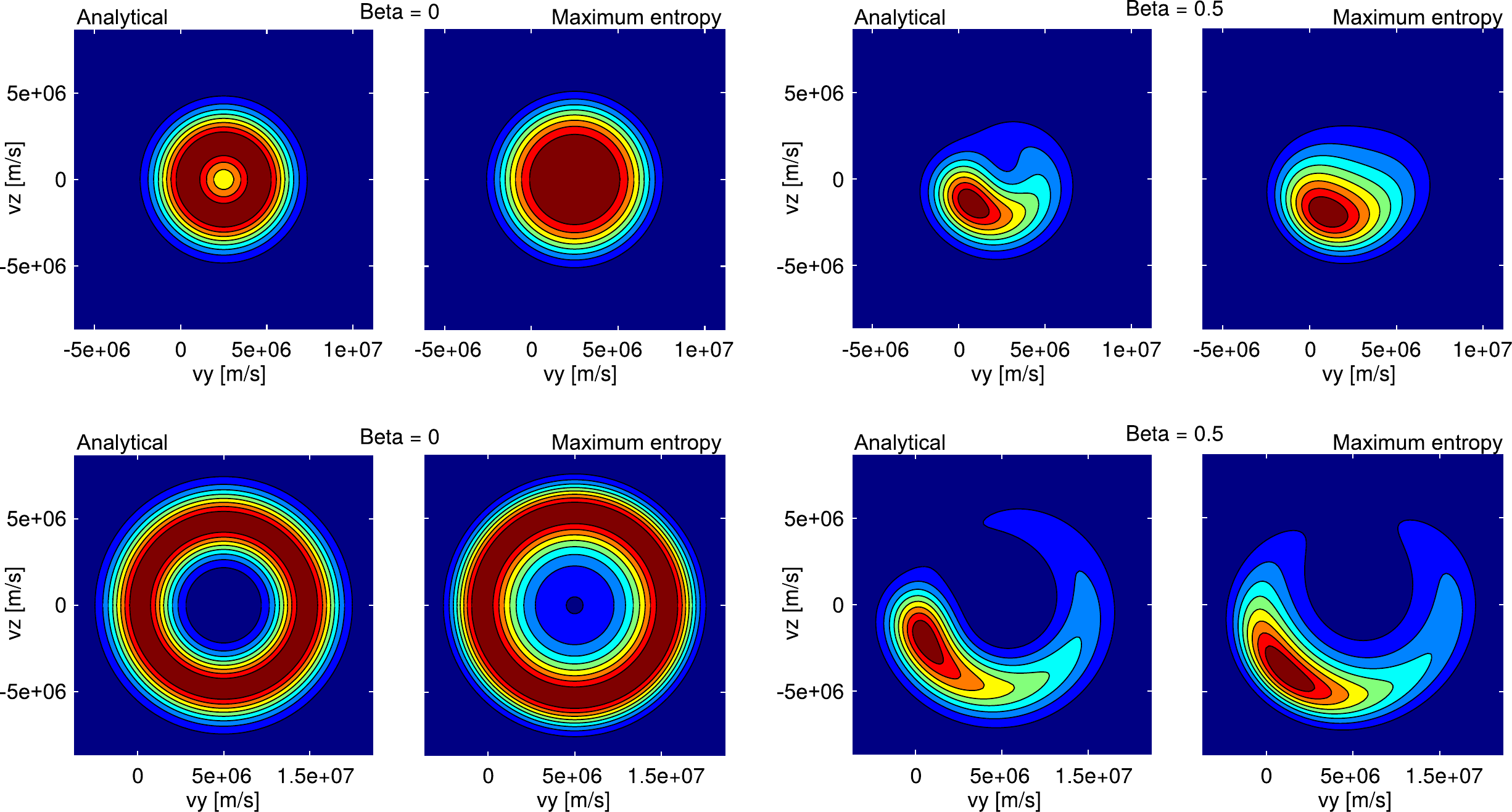}
    \caption{VDFs for different values of electric field and collisionality. Inverse Hall parameter $\beta = 0$ (left) and $\beta = 0.5$ (right). Electric field $\mathsf{E} = 25\,000$~V/m (top) and $\mathsf{E} = 50\,000$~V/m (bottom). Analytical VDFs are obtained from Shagayda \cite{shagayda2012stationary} using parameters $T_b = 100\, 000$ K (corresponding to a final temperature $T \approx 500\, 000$ K, $n_b = 10^{17} \ \mathrm{1/m^3}$, $\mathsf{B} = 0.01$ T. Maximum-entropy VDFs are obtained by numerical solution of the entropy maximisation problem.}
    \label{fig:VDF-shagayda-maxent}
\end{figure*}

The quality of the current closure is checked by feeding the first 14 moments into the entropy maximisation procedure of Eq.~(\ref{eq:minimization-problem}).
This results in the array of 14 coefficients, $\bm{\alpha}$, completely defining the VDF.
Figure~\ref{fig:VDF-shagayda-maxent} compares the analytical and maximum-entropy numerical VDFs for two different values of the drift velocity, $\mathsf{E}/\mathsf{B}$, resulting from a choice of the electric field, $\mathsf{E} = 25\,000$ V/m and $50\,000$ V/m, and by taking $\mathsf{B} = 0.01$ T.
Other parameters for the analytical VDF are chosen to be a birth temperature $T_b = 10^5$ K, characterizing the distribution of post-collision states (which results in a higher final temperature, due to the electric drift), and a number density $n = 10^{17} \ \mathrm{m^{-3}}$. 
Collisionality is expressed through the inverse Hall parameter, $\beta = \nu_c / \omega_c$, with $\nu_c$ the electron-neutral collision frequency and $\omega_c$, the cyclotron frequency.
In the original formulation, the inverse Hall parameter is directly imposed; 
the reader should refer to the original work\cite{shagayda2012stationary} for further details.
In Fig.~\ref{fig:VDF-shagayda-maxent} we compare VDFs for the cases of $\beta = 0$ and $\beta = 0.5$.
We consider the frame of reference of Fig.~\ref{fig:ref-system-thruster}, with $v_y$ the velocity component in the direction of the  $\bm{\mathsf{E}}\times\bm{\mathsf{B}}$ drift.
The VDF is plotted in the plane $(v_y, v_z)$, perpendicular to the magnetic field, as the parallel distribution function along $v_x$ results in a simple Maxwellian.
Typical conditions encountered in Hall thrusters are similar to the results in Fig.~\ref{fig:VDF-shagayda-maxent} top-left, although the magnetic field is often two or three times higher, resulting in a lower drift velocity.
The other cases show the flexibility of the current 14-moment description as the non-equilibrium nature of the
electrons is increased.
A visual analysis of the VDFs reveals that maximum-entropy distributions obtained from the iterative procedure succeed in reproducing the main features of the analytical VDFs, namely anisotropy, asymmetry and the presence of a marked hole in the center for the cases with a strongest drift.
The matching is approximated, and could be further improved by employing a higher number of moments; however,
it is felt that the degree of accuracy obtained is satisfactory, as will be shown in the following.

\begin{figure}[htpb]
    \centering
    \includegraphics[width=0.9\columnwidth]{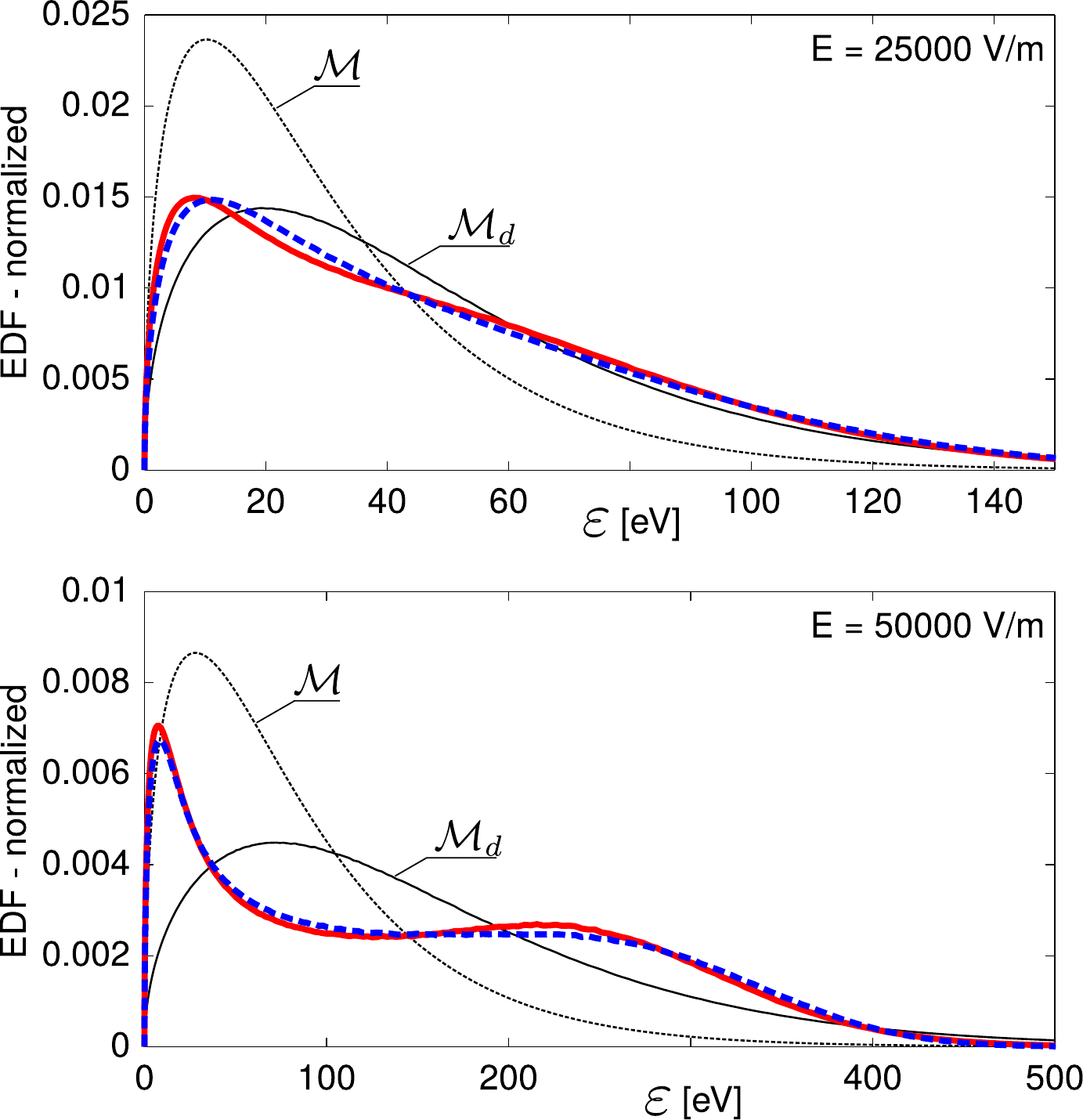}
    \caption{Normalized EEDFs obtained from Shagayda's VDFs ({\protect\redline}) [corresponding to Fedotov's solution\cite{fedotov1999electron}] and 14-moment maximum entropy closure ({\protect\bluedashedline}). For comparison, Maxwellian distribution $\mathcal{M}$ at same temperature and Maxwellian $\mathcal{M}_d$ at same temperature but drifted at velocity $\mathsf{E}/\mathsf{B}$ along $v_y$.}
    \label{fig:EDF-Shagayda-14mom}
\end{figure}

Figure~\ref{fig:EDF-Shagayda-14mom} compares the energy distribution function (EDF) for the two collision-less cases (parameter $\beta = 0$). 
Strong deviations from equilibrium are observed, especially for high values of the drift velocity (as compared to the thermal velocity), and the approximated 14-moment maximum-entropy closures is able to reproduce closely the analytical distributions.
A better matching in the EDF is expected with respect to the VDF, since the former is obtained by an integration over the velocity space.

\begin{figure}[htpb]
  \centering
  \includegraphics[width=\columnwidth]{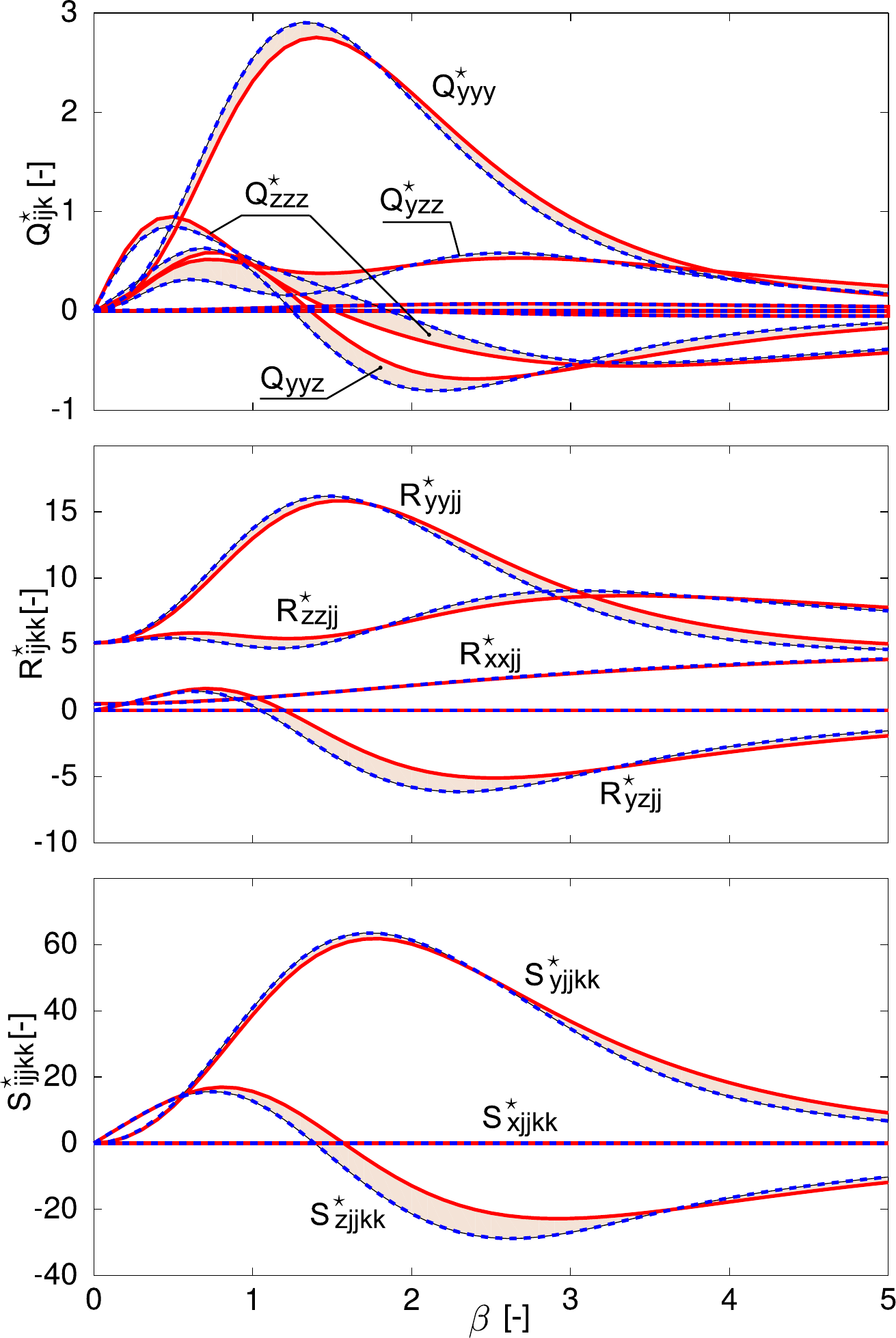}
  \caption{Non-dimensional closing moments for the case $\mathsf{E} = 50\,000 \, \mathrm{V/m}$. Moments from Shagayda's VDF ({\protect\redline}) and from the iterative solution of the 14-moment entropy-maximisation problem ({\protect\bluedashedline}). The mismatch between the lines has been highlighted for clearer identification.}
  \label{fig:Q-R-S-50000}
\end{figure}

Since the ultimate goal is the 14-moment fluid-like formulation of the problem, rather than the distribution function itself, we are more interested in its moments, especially the value of 
the closing moments $Q_{ijk}$, $R_{ijkk}$, and $S_{ijjkk}$.
Such moments have been computed from the analytical and approximated VDFs for different values of the parameter $\beta$ and are compared in Fig.~\ref{fig:Q-R-S-50000} for the case of $\mathsf{E} = 50\,000$ V/m.
For an easier representation, their non-dimensional value is shown, by dividing the moment by the density $\rho$ and by powers of the characteristic thermal velocity, $\sqrt{P/\rho}$, giving
\begin{subequations}\label{eq:nondim-Q-R-S}
\begin{align}
    Q^\star_{ijk} &= \frac{Q_{ijk}}{\rho (P/\rho)^{3/2}} \, , \\
    R^\star_{ijkk} &= \frac{R_{ijkk}}{\rho (P/\rho)^{4/2}} \, , \\ 
    S^\star_{ijjkk} &= \frac{S_{ijjkk}}{\rho (P/\rho)^{5/2}} \, .
\end{align}
\end{subequations}

\noindent It shall be stressed that in such comparison, the first 14 moments are exactly the same for the analytical and the maximum-entropy schemes, since they are enforced by the optimization algorithm.
The density and pressure in Eqs.~\ref{eq:nondim-Q-R-S} are therefore known for each case.
Fig.~\ref{fig:Q-R-S-50000} shows a good matching between the analytical and the approximated maximum-entropy closing moments. 
Conditions typical of Hall thrusters are located around values of $\beta \ll 1$ (high Hall parameter), where the matching proves very good.

While the present results are obtained with the classical iterative maximum-entropy closure, the approximated interpolative closure of McDonald and Torrilhon \cite{mcdonald2013affordable} was also applied to these test cases, showing fully comparable levels of accuracy.
The same analysis was performed for the case with $\mathsf{E} = 25\,000$ V/m, leading to completely analogous results.
It is important to remark that an exact match between analytical and approximated closing moments is most likely not needed in a real-life scenario, since the effect of closing moments is mitigated to some degree by the presence of many other lower order moments inside the fluxes.

\begin{figure}[htpb]
  \centering
  \includegraphics[width=\columnwidth]{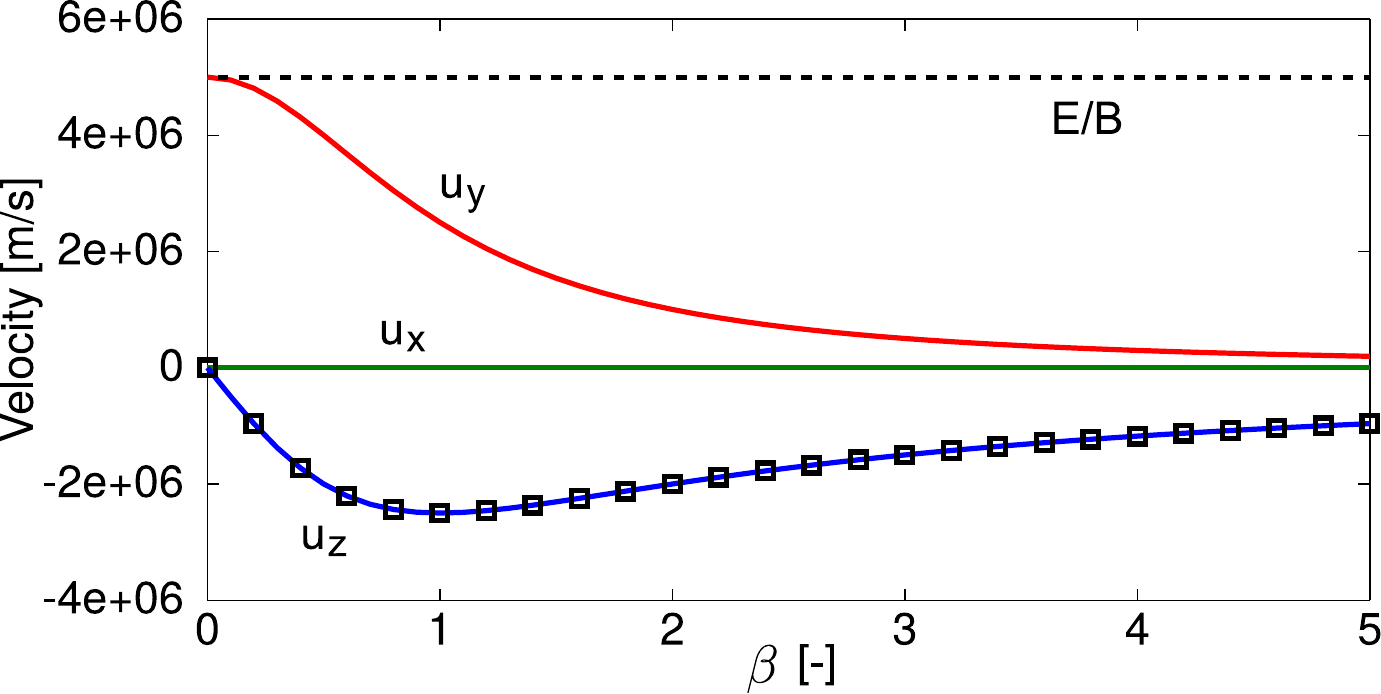}
  \caption{Solid lines: velocity components for the case $\mathsf{E} = 50\,000$ V/m. Symbols: cross-field trasport from Eq.~(\ref{eq:cross-field-trasport}).}
  \label{fig:velocity-xyz-vs-beta}
\end{figure}

Finally, it should be noted that, for large values of $\beta$, collisions become dominant and transform the distribution function into a Maxwellian, and all moments eventually reach their characteristic values at equilibrium, for $\beta \rightarrow \infty$: 
odd-order moments vanish, the pressure tensor becomes diagonal and isotropic, the fourth order central moment $R_{iijj}$ reaches the value $15 P^2/\rho$ and the closing moments, $R_{ijkk}$, tend to $5P^2\delta_{ij}/\rho$.
Some further interpretation can be drawn by observing the average velocity, in Fig.~\ref{fig:velocity-xyz-vs-beta}.
For $\beta = 0$, we have a collision-less formulation with the velocity component $u_y$ exactly equal to $\mathsf{E}/\mathsf{B}$ and the cross-field velocity $u_z$ equal to zero.
At higher values of $\beta$, $u_y$ gradually decreases and $u_z$ follows the classical result for cross-field diffusion,
\begin{equation}\label{eq:cross-field-trasport}
    u_z = \frac{e/(m \nu_c)}{\omega_c^2/\nu_c^2 + 1} \mathsf{E} \, .
\end{equation}

\noindent with $e$ the elementary  electric charge.
Such an accurate matching for the cross-field mobility may seem surprising for non-equilibrium conditions, but arises from the assumption of artificially imposing the collision frequency, and neglecting its dependence on the actual state of the electron gas.
Moreover, the presence of space gradients can also be expected to reduce this degree of accuracy.

\section{Electric and magnetic field source terms}\label{sec:EM-fields-sources}

As Section~\ref{sec:analytical-vs-14mom} showed that the 14-moment system is able to reproduce the required distributions, the next step consists in formulating the source terms for the dynamical system in Eq.~(\ref{eq:governing-equations-fluid}).

Electric and magnetic source terms, $\bm{S}_{em}$, for the 14 moments are obtained by computing the averages in Eq.~(\ref{eq:generalized-moment-eq}) involving the velocity derivatives of the generating functions $\phi(v)$,
\begin{equation}
    {S}_{em} 
    = \frac{q}{m}\left< \bm{\mathsf{E}} \cdot \frac{\partial \phi}{\partial \bm{v}} \right>
    + \frac{q}{m}\left< \left( \bm{v} \times \bm{\mathsf{B}} \right) \cdot \frac{\partial \phi}{\partial \bm{v}} \right> \, .
\end{equation}

We assume for simplicity that $\bm{\mathsf{B}} = \mathsf{B} \hat{x}$ and $\bm{\mathsf{E}} = \mathsf{E} \hat{z}$, as shown in Fig.~\ref{fig:ref-system-thruster}.
Extending the results to fields in arbitrary directions only requires one to repeat the present calculations for the additional components.
As an example, we consider the computation of the source term for the $z$-direction momentum, obtained from the choice $\psi = m v_z$ and corresponding to the conserved variable $U_5$.
The gradient of $\phi(v)$ results in $\partial \phi / \partial \bm{v} = (0, 0, m)$, hence
\begin{equation}
    S_{em}^{(5)} = \frac{q}{m} \left[ \mathsf{E}\left<   m \right> - \mathsf{B} \left< v_y  m \right> \right] = \frac{q}{m} \rho \, \mathsf{E} - \frac{q}{m} \rho u_y \, \mathsf{B} \, ,
\end{equation}

\noindent where we recognize the conserved moments $U_1 \equiv \rho$ and $U_3 = \rho u_y$, defined in Eq.~(\ref{eq:Ui-definitions}).
The computation for all other terms is analogous and gives
\begin{equation}
  \bm{S}_{em} 
= \frac{q}{m}
  \begin{pmatrix}
    0 \\                               
    0 \\                    
    \mathsf{B} \, U_4 \\ 
    \mathsf{E} \, U_1 - \mathsf{B} \, U_3\\                               
    0 \\ 
    \mathsf{B} \, U_7 \\ 
    \mathsf{E} \, U_2 - \mathsf{B} \, U_6 \\       
    2 \mathsf{B} \, U_9 \\       
    \mathsf{E}  \, U_3 + \mathsf{B} \, \left( U_{10} - U_8 \right)\\         
    2\mathsf{E}  \, U_4 - 2 \mathsf{B} \, U_9 \\                                       
    2\mathsf{E} \, U_7 \\      
    2\mathsf{E}  \, U_9  + \mathsf{B} \, U_{13}\\    
    \mathsf{E} \, \left( U_5 + U_8 + 3 U_{10}\right) - \mathsf{B} \, U_{12}\\                     
    4\mathsf{E} \, U_{13}                    
  \end{pmatrix} \, .
\end{equation}

\noindent Note that this source does not introduce any additional moments to the problem, such that it can be computed exactly, regardless of the particular closure employed.
For every moment equation of order $p$, the electric term requires a moment of order $p-1$.  This is due to the derivative in the source term.
On the other hand, the magnetic terms involve moments of the same order as the equation itself, due to the cross product with the velocity.
The effect is a mixing of the components in the perpendicular field.
In the present reference system, the magnetic field does not affect the $v_x$ velocity component, therefore it has no effect in the moments $U_{2}=\left<mv_x\right> = \rho u_x$, $U_{5}=\left<mv_xv_x\right> = \rho u_x^2 +P_{xx}$ and $U_{11}=\left<mv_xv^2\right>$.
Also, as the fourth order moment $U_{14}=\left<mv^4\right>$ is a contraction over the three directions and is strictly related to the square of the energy, and is thus not affected by the magnetic field.

Electric and magnetic fields typically depend on the solution itself through the charge density and current appearing in the Maxwell equations. 
However, in the case of externally imposed fields (or considering the fields at a given timestep, in the framework of an explicit time-integration method for example), this term becomes linear in the conserved variables, and can be rewritten as $\bm{S}_{em} = \matr{A} \bm{U}$.
This can be of particular use in building time integration schemes for the solution of the moment
equations.  Indeed, considering space-homogeneous conditions and neglecting collisions, the system becomes
\begin{equation}
    \frac{{\rm d} \bm{U}}{{\rm d} t} = \matr{A} \, \bm{U} \, ,
\end{equation}

\noindent whose solution is
\begin{equation}
    \bm{U}(t) = \exp\left[\matr{A} (t - t_0)\right] \bm{U}(t_0) \, .
\end{equation}

\noindent This can be the basis for the formulation of time integration schemes that reproduce consistently the effect of electro-magnetic fields on all moments.
While this could be of little importance in problems when the external work done on the system is large, the simulation of a closed system for long times requires particular care on the choice of time integrators, to avoid artificial phenomena such as the appearance of negative temperatures.


\section{BGK-like electron-neutral collisions in low-temperature plasmas}\label{sec:collision-term}

The focus of this work is the investigation of the 14-moment system, therefore a detailed or very accurate description of collisional processes is not a priority at this stage and we consider only BGK-like collision operators.\cite{bhatnagar1954model}
We only consider collisions of hot electrons with a cold background gas;
electron-electron collisions could be easily introduced, and would only increase the accuracy of the 14-moment system by providing additional paths towards local equilibrium distribution functions.
In a sense, if the system works well in the present work, it will most likely work also in conditions with more collisions.

The success of the approximated BGK collision operator in rarefied gas dynamics is to be attributed to its great simplicity and the possibility of a trivial derivation of source terms for moment equations.
Such operator automatically recovers Boltzmann's H-theorem by equilibrating the gas towards a local Maxwellian.
This operator has been extended in a number of scenarios, including multi-temperature monoatomic and polyatomic gases and chemically reacting mixtures.\cite{burgers1969flow,bisi2016bgk,bernard2019bgk}
This type of source-term approximation requires some further adaptation if a reasonable description of electron-neutral collisions in low-temperature non-thermal plasmas is sought.

First, one should consider that low-temperature plasmas are characterized by low translational temperature for heavy species and an electron temperature that is much more elevated.
In most situations, the steady-state reached by electrons is a non-thermal condition arising from a balance between energy lost in collisions with cold background species and external energy supplied by the electric field.
Aiming at retrieving the H-theorem in the classical multi-species gas dynamic sense (relaxation of electrons and heavy species at comparable temperatures) may be desirable for the sake of describing processes such as the relaxation behind a strong shock wave or for \textit{thermal} plasmas, but is probably of little interest for the description of the present problem.
Therefore, we formulate the collision operator so as to represent a system in sustained non-thermal conditions.

Secondly, for particles with large mass disparity, the energy exchange due to elastic collisions becomes extremely ineffective, and each particle roughly conserves its energy.
Indeed, denoting by $m$ the mass of the light particle, $M$ the heavy species mass, $\chi$ the angle at which the particle velocities are rotated by the collision, and by $\Delta \varepsilon / \varepsilon$ the relative exchange of energy, in the limit of $m \ll M$ one finds\cite{lamarsh1966introduction}
\begin{equation}\label{eq:light-heavy-energy-exchange}
  \frac{\Delta \varepsilon}{\varepsilon} = 2 \frac{m}{M} \left( 1 - \cos \chi \right) \, .
\end{equation}

\noindent For electrons and xenon neutrals, the mass ratio is roughly $m/M \approx 10^{-5}$, such that the initial energy of a hot electron will be lost, on average, after some $10^5$ elastic collisions.
In the following, we will only consider the case of isotropic scattering, which is a good approximation only for relatively low collision energies.\cite{vahedi1995monte,surendra1990self}

Section \ref{sec:BGK-electron-neutral} introduces a simple Maxwellian relaxation model based on the previous considerations; 
in Section \ref{sec:isotropic-E-preserving-collis} we introduce a different isotropic collision model which aims at representing more closely the effect of high mass disparity; in Section \ref{sec:energy-loss-collisions} the energy exchange with background neutrals is introduced, together with excitation and ionization collisions,
and in Section \ref{sec:collis-frequency} the expression of the collision frequency is discussed for non-equilibrium distributions.


\subsection{BGK-like Maxwellian relaxation}\label{sec:BGK-electron-neutral}

We obtain a first model for eleastic collisions by assuming that the electron-neutral collision operator can be approximated by
\begin{equation}
    \mathscr{C} = - \nu \left(f - \mathcal{M}\right) \, ,
\end{equation}

\noindent where the Maxwellian distribution, $\mathcal{M}$, represent the distribution of post-collision states of electrons.
This model resembles the single-species BGK formulation, but differs in that:

\begin{itemize} 
  \item Neutrals are cold and have low bulk velocity with respect to the colliding electrons;
  \item The exchange of energy $\Delta \varepsilon / \varepsilon$ between the hot electron and the massive neutral is neglected;
  \item The scattering is assumed isotropic in the center of mass frame, coinciding with the heavy neutral.
\end{itemize}
  
\noindent Therefore, the average velocity of the post-collision Maxwellian is zero and the post-collision electron temperature, $T_p$, is obtained as\cite{shagayda2012stationary}
\begin{equation}\label{eq:temperature-post-collision}
    \tfrac{3}{2} k_B T_{p} = \tfrac{3}{2} k_B T + \tfrac{1}{2} m u^2 \, .
\end{equation}

\noindent The corresponding pressure is $P_p = n k_B T_p$, with $n$ the local electrons number density (unchanged by the collision process since only elastic collisions are considered) and $k_B$ the Boltzmann constant.
The collisional source for the 14 moments is obtained from the vector of generating weights $\bm{\Phi}$ as
\begin{equation}
  \mathbf{S}_c^{\mathcal{M}}
  = - \left< \bm{\Phi} \nu \left( f - \mathcal{M} \right) \right>
  = - \nu \left[ \left< \bm{\Phi} f \right> - \left< \bm{\Phi} \mathcal{M} \right> \right] \, ,
\end{equation}

\noindent which reduces to the difference between the moments of $\phi$ in the current state (distribution $f$) and their value for a Maxwellian at temperature, $T_p$, and zero velocity.
Due to the symmetry of the Maxwellian, all central moments of odd order are zero and the source terms are easily evaluated from the following criterium:
\begin{itemize}
    \item All odd-order (in the velocity components) moment equations relax to zero;
    \item Even-order moments relax to the value for a Maxwellian at the post-collision temperature $T_p$.
\end{itemize}

\noindent This can be verified from a direct calculation of the integrals, and results in
\begin{equation}\label{eq:BGK-source-term}
    \bm{S}_c^{\mathcal{M}} = 
    \begin{pmatrix}
       0 \\
       - U_2/\tau \\
       - U_3/\tau \\
       - U_4/\tau \\
       - (U_5 - P_p)/\tau \\       
       - U_6/\tau \\
       - U_7/\tau \\
       - (U_8 - P_p)/\tau \\
       - U_9/\tau \\
       - (U_{10} - P_p)/\tau \\
       - U_{11}/\tau \\
       - U_{12}/\tau \\
       - U_{13}/\tau \\
       - (U_{14} - 15P_p^{2}/\rho)/\tau
    \end{pmatrix} \, ,
\end{equation}

\noindent where the moments $U_i$ are defined in Eq.~(\ref{eq:Ui-definitions}) and the characteristic time $\tau$ is the inverse of the electron-neutral collision frequency, discussed in Section \ref{sec:collis-frequency}.

The simplicity of this collision model makes it attractive for approximated calculations or for situations of low collisionality. 
There are however some important drawbacks.
First, this collision model implies that all moments relax towards local equilibrium at the same rate, which is a known issue of most BGK approximations, whenever the collision frequency does not depend on the microscopic velocity.\cite{struchtrup2005macroscopic}
Moreover, for collisions between hot electrons and background particles that are colder and more massive, the idea of relaxing towards a Maxwellian is also questionable---a generalization is discussed in the next section.


\subsection{BGK-like model for large mass disparity}\label{sec:isotropic-E-preserving-collis}

As shown in Eq.~(\ref{eq:light-heavy-energy-exchange}), in elastic collisions with large mass disparity the energy exchange becomes extremely ineffective, and each particle roughly conserves its energy during a collision.
To a first approximation, a collision has the effect of rotating the relative velocity of a particle without changing its energy.
To illustrate this idea, one can consider an initial beam of electrons impacting a target, represented by a Dirac delta in velocity space, the randomizing effect of collisions will gradually transform the distribution into a void sphere, all the electrons would remain concentrated at its crust, since the energy of each electron is conserved.
The same argument can be extended to any distribution function, considered as a ``sum of delta functions'':
whereas the distribution of velocities evolves towards isotropy for effect of the collisions, the distribution of energies would not be affected.
Therefore, the BGK assumption of relaxation towards a Maxwellian is reasonable for colliding pairs with comparable mass, but not for particles with large mass disparity.
From the previous observation we build a BGK-like collision operator in the form
\begin{equation}
  \mathscr{C} = - \nu ( f - f^{\mathrm{iso}} ) \, ,
\end{equation}

\noindent where $f^{\mathrm{iso}}$ is an isotropic distribution with the same energy content as
$f$, such that the condition,
\begin{equation}
  f^{\mathrm{iso}} (\varepsilon) \equiv f(\varepsilon)  \ \ \implies \ \ f^{\mathrm{iso}} (v^2) \equiv f(v^2) \, ,
\end{equation}

\noindent holds. The actual shape of $f^{\mathrm{iso}}$ in the velocity does not matter for the sake of computing the 14 moments, but the following considerations will suffice:
first, we consider that the isotropic distribution $f^{\mathrm{iso}}$ is symmetric in the velocities, such that its odd-order central moments are all zero.
The average velocity of $f^{\mathrm{iso}}$ is also zero, as we are considering isotropic scattering and the bulk velocity of target neutrals is assumed negligible.
Odd moments will therefore relax towards zero, as in the Maxwellian relaxation model.
The temperature of the post-collision isotropic distribution is also obtained from Eq.~(\ref{eq:temperature-post-collision}).

The only remaining quantity is the fourth-order moment $U_{14} = \left<m v^4\right>$.
Intuitively, since the energy of each electron is conserved by the collision, its square will also be conserved, such that the global moment $\left<m v^4\right>$ will not change, being the average of that quantity over all individual particles.
This can be formally retrieved by writing the definition of the post-collision fourth moment in spherical coordinates:
\begin{equation}
  \left< m v^4 \right>^{\mathrm{iso}} \equiv \int m v^4 f^{\mathrm{iso}}(v^2) \, v^2 \sin \theta \, \mathrm{d} v \, \mathrm{d} \psi \, \mathrm{d} \theta \, ,
\end{equation}

\noindent where we highlight the dependence on the velocity \textit{modulus} by writing $v^2$.
Since $f^{\mathrm{iso}}(v^2) \equiv f(v^2)$, we have
\begin{equation}
  \left< m v^4 \right>^{\mathrm{iso}} = \left< m v^4 \right> \, .
\end{equation}

\noindent Therefore, this collision operator has no effect on the contracted fourth-order moment, since it preserves the EEDF.
The energy-preserving operator thus generates a collision source term almost equal to the Maxwellian relaxation source of Eq.~(\ref{eq:BGK-source-term}), but with the last term equal to zero:
\begin{equation}
    \bm{S}^{\mathrm{iso}}_c(1,\cdots\!,13) = \bm{S}_c^{\mathcal{M}}(1,\cdots\!,13) \ \ \mathrm{and} \ \ \bm{S}^{\mathrm{iso}}_c(14) = 0 \, .
\end{equation}


\subsection{Energy loss and inelastic collisions}\label{sec:energy-loss-collisions}

One can generalize the applicability of the collision model by including the energy lost by hot electrons due to the effects of (i) elastic collisions, (ii) electronic excitation of neutrals, and (iii) ionization.

\subsubsection*{Elastic collisions}
The average energy lost during an elastic collision with a heavy particle can be expressed by averaging Eq.~(\ref{eq:light-heavy-energy-exchange}) over the possible deflection angles, $\chi$.
The angle, $\chi$, requires that the differential cross-section be considered, which is, in general, a function of both the impact parameter and the collision energy.
In the present case, we consider isotropic scattering for simplicity, limiting the validity of our model to relatively low collision energies.\cite{vahedi1995monte}
From this assumption, when averaged over the population of electrons, the energy lost results in $\left< \Delta \varepsilon / \varepsilon\right>_\chi = 2 m/M$, where we denoted by $\left< \, \right>_\chi$ the average over the distribution of deflection angles.
Indeed, the quantity $(1 - \cos\chi)$ can be shown to be uniformly distributed\cite{lamarsh1966introduction} over the interval $[0,2]$, with an average value of $1$.
The post-collision distribution (either Maxwellian or general isotropic) is thus characterized by an energy reduced by such factor, and the post-collision temperature results:
\begin{equation}\label{eq:temperature-post-collision-lossenergy}
    \tfrac{3}{2} k_B T_p = \left( \tfrac{3}{2} k_B T + \tfrac{1}{2}m u^2\right)\left( 1 - 2 m/M \right) \, .
\end{equation}

\noindent By assuming isotropic scattering, the distribution of post-collision states will still be symmetric, such that its central odd-order moments are still zero, and the bulk velocity is also zero.
The pressure components relax to $P_p = n k_B T_p$ as for the other models, with $T_p$ from Eq.~(\ref{eq:temperature-post-collision-lossenergy}).
The Maxwellian relaxation model of Eq.~(\ref{eq:BGK-source-term}) is completely defined at this point.

For the general isotropic-collision model of Section~\ref{sec:isotropic-E-preserving-collis} we also need to investigate the fourth-order moment.
Following Eq.~(\ref{eq:light-heavy-energy-exchange}), we can compute the post-collision energy, $m v_i^{\prime 2}/2$, of an electron ``$i$'' deflected by the angle, $\chi$.
By squaring the energy, we can write 
\begin{multline}
  m^2 v_i^{\prime\ 4} 
= m^2 v_i^4 \left( 1 - \tfrac{2 m}{M} (1 - \cos \chi) \right)^2 \\
= m^2 v_i^4 \left( 1 + \tfrac{4 m^2}{M^2} (1 - \cos \chi)^2 - \tfrac{4 m}{M} (1 - \cos \chi) \right) \, .
\end{multline}

\noindent The fourth-order moment is obtained by averaging this term over all possible deflection angles.
As mentioned above, the quantity $(1-\cos\chi)$ is uniformly distributed in the case of isotropic collisions.
Since $(1 - \cos\chi)^2$ is the square of a uniformly distributed quantity, its probability distribution is also known,\cite{stirzaker2003elementary} and its average can be found to take the value $4/3$.
Therefore, considering all possible deflection angles, a collision \textit{reduces}, in average, the quantity $v^{\prime\,4}$ by the factor
\begin{equation}
  v^{\prime\,4} = v^4 \left( 1 + \frac{16}{3}\frac{m^2}{M^2} - \frac{4m}{M} \right) \, .
\end{equation}

\noindent At this point we are in a position to compute the fourth-order moment as an average of the quantity $mv^4$ over the distribution function:
\begin{equation}
  \left< m v^4 \right>^{\mathrm{iso}} = \left( 1 + \frac{16}{3} \frac{m^2}{M^2} - \frac{4 m}{M} \right) \left< m v^4 \right> \, .
\end{equation}

\noindent again, in the case of $m/M \rightarrow 0$, the conservation is restored.
The collision source term for the general isotropic model will be exactly like the Maxwellian source for the first 13 moments, using the proper reduced pressure from Eq.~(\ref{eq:temperature-post-collision-lossenergy}), and the last entry will be
\begin{equation}
    \bm{S}^{\mathrm{iso}}_c(14) = +\nu U_{14} \left( \frac{16}{3} \frac{m^2}{M^2} - \frac{4 m}{M} \right) \, .
\end{equation}
  
\noindent This source term is slightly negative, since $m \ll M$, and therefore reduces slightly the fourth order moment in time.
In the limit of infinite time, if no energy is supplied to the system, this collision operator would result in a progressive cooling down of electrons. 
However, as anticipated, this model is targeted for representing steady states of non-thermal plasmas where electrons are maintained at high temperature with respect to the background neutrals.
More complete models that relax towards the temperature of background neutrals could be derived following Burgers.\cite{burgers1969flow}

\subsubsection*{Excitation and ionization}

The present model can be supplemented by inelastic collisions by including more source terms introducing excitation and ionization processes.
If a Maxwellian relaxation term is assumed, the sources for the 14 moment equations result in a sum of different terms for the different processes considered.
Each term will be in the form of Eq.~(\ref{eq:BGK-source-term}), but with post-collisions temperatures/pressures reduced by the energy lost in the considered process.\cite{shagayda2012stationary}
Finally, if ionization reactions are considerend, a difference in the number density for the pre-collision and post-collision distributions is to be included, and the post-collision energy is to be split between the primary and secondary electrons, resulting in a further cooling down of the electrons.\cite{vahedi1995monte}


\subsection{Collision frequency for non-equilibrium distributions}\label{sec:collis-frequency}

The collision frequency of one electron with a background of target molecules of density $n_{BG}$ can be expressed as $\nu_c = n_{BG} \left< \sigma(v_r) \, v_r \right>_r$, where $v_r$ is the relative velocity between the considered electron and the population of target species, and $\left< \bullet \right>_r$ denotes the integration over the distribution of relative velocities.\cite{kuppermann1968chemical}
In classical gas dynamics, the collision frequency is often expressed by assuming a Maxwellian distribution of relative velocities, resulting in $\nu_c = n \, \sigma \, v^{th}$, with $v^{th}=(8 k_B T/(\pi m))^{1/2}$ the thermal velocity.

In low-temperature plasmas, electrons have much higher thermal and drift velocity than the slow and cold background neutrals, such that the relative velocity coincides with the absolute velocity of electrons, and the distribution of relative velocities is effectively the (normalized) distribution function of electrons in the lab frame.
Further assuming a constant cross-section $\bar{\sigma}$ for simplicity, we have:
\begin{equation}\label{eq:cross-section-formula}
    \nu_c = n_{BG} \ \bar{\sigma} \left[ \frac{1}{n} \int_0^{+\infty} \!\!\! v f(v) \, \mathrm{d}v \right] = n_{BG} \ \bar{\sigma} \left< |v| \right> \, .
\end{equation}

\begin{figure}[htpb]
  \includegraphics[width=\columnwidth]{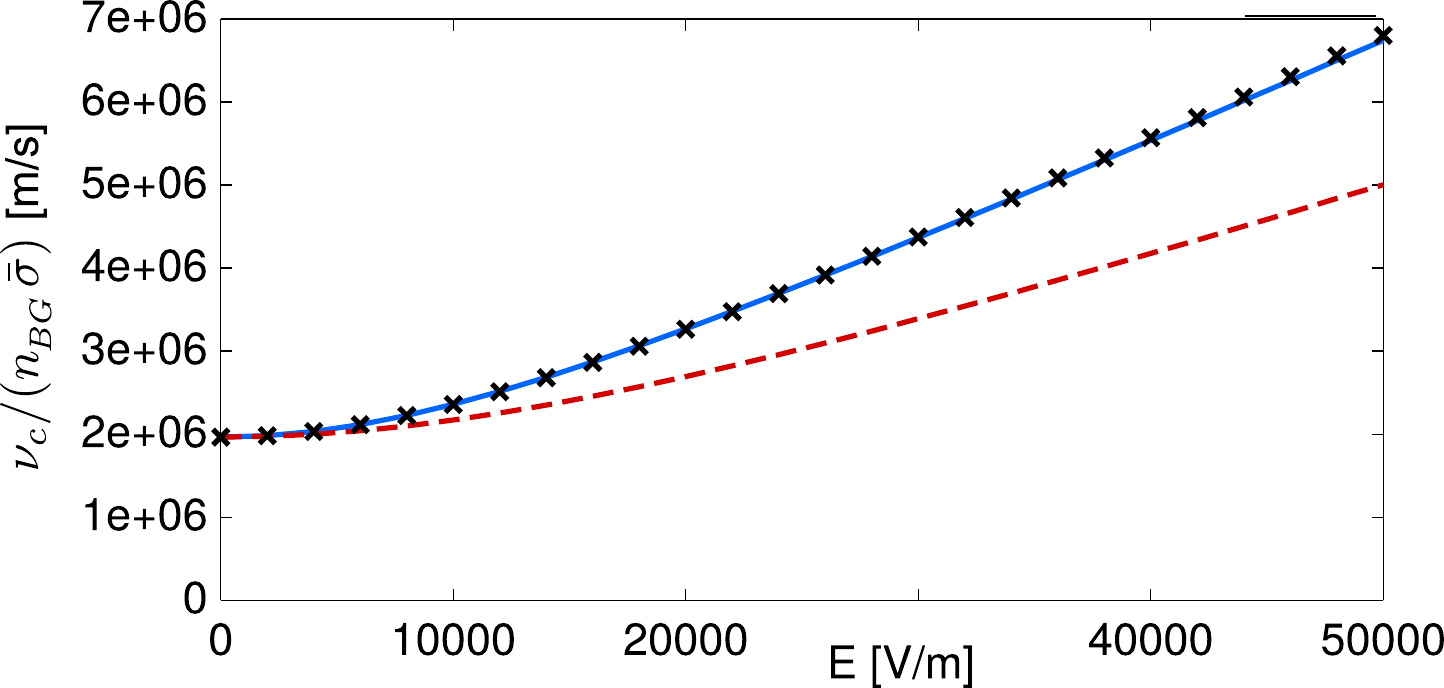}
  \caption{Reduced collision frequency for the distributions of Section \ref{sec:analytical-vs-14mom} with $\beta = 0$, $T_b = 10^5$ K and $\mathsf{B} = 0.01$ T, for various electric fields.
  For $\mathsf{E} = 0$ V/m, the distribution is a Maxwellian with zero average velocity. For large drift velocities $u_d = \mathsf{E}/\mathsf{B}$ the distribution is strongly out of equilibrium.
  Numerical integration of the non-equilibrium distribution ({\protect\myblueline}); thermal energy formula ({\protect\myreddashedline}) and approximation with total energy of Eq.~(\ref{eq:coll-freq-approx}) (symbols).}
  \label{fig:coll-freq-adim-approx}
\end{figure}

\noindent Therefore, $\nu_c$ depends on the degree of non-equilibrium of $f$.
Even more importantly, in strongly magnetized conditions, the bulk velocity of $f$ (due to the $\mathsf{E}/\mathsf{B}$ drift and cross-field transport) can be comparable with the thermal velocities, as can be clearly seen in Figs.~\ref{fig:VDF-shagayda-maxent} and $\ref{fig:velocity-xyz-vs-beta}$.
In the conditions of this work, rather than non-equilibrium, the non-zero bulk velocity appears to be the leading effect in determining the collision frequency, and can drive large discrepancies from the actual collision frequency, as shown in Fig.~\ref{fig:coll-freq-adim-approx}.
To obtain a more reliable result one could approximate the collision frequency by the value obtained from a drifted Maxwellian.

We propose an alternative approach, consisting in computing the average velocity as $\left< |v| \right> \approx \sqrt{8 k_B T^{\mathrm{tot}}/(\pi m)}$, where the temperature $T^{\mathrm{tot}}$ accounts for the \textit{total} kinetic energy rather than only the thermal contribution: $T^{\mathrm{tot}} = 2 n \varepsilon^{\mathrm{tot}} / (3 k_B)$, with $\varepsilon^{\mathrm{tot}}$ the electrons total energy per unit mass (ordered kinetic plus thermal).
The resulting collision frequency is necessarily higher than the one based on thermal motion only and reads
\begin{equation}\label{eq:coll-freq-approx}
    \nu_c = n_{BG} \ \bar{\sigma} \ \sqrt{\frac{16 \, \varepsilon^{\mathrm{tot}}}{3 \pi}} \, .
\end{equation}

\noindent Fig.~\ref{fig:coll-freq-adim-approx} compares this approximation to the actual value for the non-equilibrium collisionless distributions of Section~\ref{sec:analytical-vs-14mom} (parameter $\beta = 0$), showing high accuracy for the considered conditions.
The same analysis was repeated for collisional distributions ($\beta > 0$), showing analogous reliability for the approximated formula.
Detailing the treatment of energy-dependent cross-sections could be very important for obtaining accurate and reliable simulations, however this lies beyond the scope of this work.
As a first approximation, we propose to evaluate them using the approximated value of average velocity $\left< |v| \right>$ discussed above.

\section{Test case: homogeneous relaxation}\label{sec:0D-relaxation}

In Section~\ref{sec:analytical-vs-14mom}, the 14-moment approximation is compared to analytical distributions, providing a first verification of the suitability of such a description.
In this Section, we propose a further verification for the 14-moment closure, and simulate the space-homogeneous relaxation of electrons in a background of cold and slow neutrals.
Such a test case includes all the ingredients that have been developed in the previous sections:
the electro-magnetic field sources and electron-neutral collisions.
This test case is thus propaedeutic to the future application of the 14-moment closure in multiple
space dimensions.

We consider an initial anisotropic and drifted distribution for the electrons velocities, described by the Gaussian
\begin{equation}
\begin{cases}
    f_0 = A \ \exp\left[- \frac{m}{2 k_B} \left( \frac{v_x^2}{T_{0x}} - \frac{(v_y - u_{0y})^2}{T_{0y}} - \frac{v_z^2}{T_{0z}}\right)\right] \\
    A = n_0 \left(\frac{m}{2 \pi k_B} \right)^{3/2} \left(\frac{1}{T_{0x} T_{0y} T_{0z}} \right)^{1/2} \, ,
\end{cases}
\end{equation}

\noindent with temperatures $T_{0x} = 10\,000$, $T_{0y} = 20\,000$ and $T_{0z} = 5000$~K, non-zero bulk velocity $u_{0y} = 30\,000$~m/s and density $n_0 = 10^{17} \,\mathrm{m^{-3}}$.
Electric and magnetic fields are switched on at time $t = 0$ to the constant value of $\bm{\mathsf{E}} = 20\, 000 \, \hat{\bm{z}}$~V/m and $\bm{\mathsf{B}} = 0.02 \, \hat{\bm{x}}$~T.
Electro-magnetic quantities are chosen as to be representative of Hall thruster devices,\cite{ahedo2001one} and the initial thermodynamic state so as to present some degree of anisotropy in order to stress the model.

Any reasonable physical model for this problem would consider at least three time-scales:
\begin{enumerate}
  \item The cyclotron frequency, at which the VDF spins around the magnetic field and its moments oscillate;
  \item The collision frequency, at which the VDF relaxes towards a somehow isotropic distribution and the electrons velocity adapts to the background velocity;
  \item The time-scale for energy loss, orders of magnitude longer than the collision frequency, due to the strong mass disparity.
\end{enumerate}

Since the scales are very different, the problem is not trivial and requires some computational efforts, especially from the kinetic perspective.
Moreover, if one wishes to obtain steady states with reasonable values for the temperature, then excitation and ionization reactions should be accounted for, as neglecting them would result in unphysically large Ohmic heating.
This additionally requires the reaction cross-sections to be considered, and a strategy for scaling the simulated particles, since an exponential growth is eventually expected.

In this section we consider a toy-model: 
the collision operator is approximated by the BGK-like model of Section~\ref{sec:BGK-electron-neutral}, where the temperature of the post-collision Maxwellian is not taken from the local energy of electrons, but is fixed at a value $T_p = 10\,000$ K.
The collision rate $\nu_c$ is taken as a fraction of the cyclotron frequency such that $\beta = \nu_c/\omega_c = 0.3$. 
In this way, the Hall parameter is fixed and the long time scale associated to the energy relaxation is removed from the problem.
Considering the previous assumptions, the kinetic equation becomes
\begin{multline}
    \frac{\partial f}{\partial t} 
    + \frac{q \mathsf{E}}{m} \frac{\partial f}{\partial v_z} 
    + \frac{q \mathsf{B}}{m} \left[ v_z \frac{\partial f}{\partial v_y} - v_y \frac{\partial f}{\partial v_z}\right] \\
    = - \beta \, \omega_c \left[ f - \mathcal{M}(T_p) \right] \, .
\end{multline}

\begin{figure*}[htpb]
    \centering
    \includegraphics[width=1.0\textwidth]{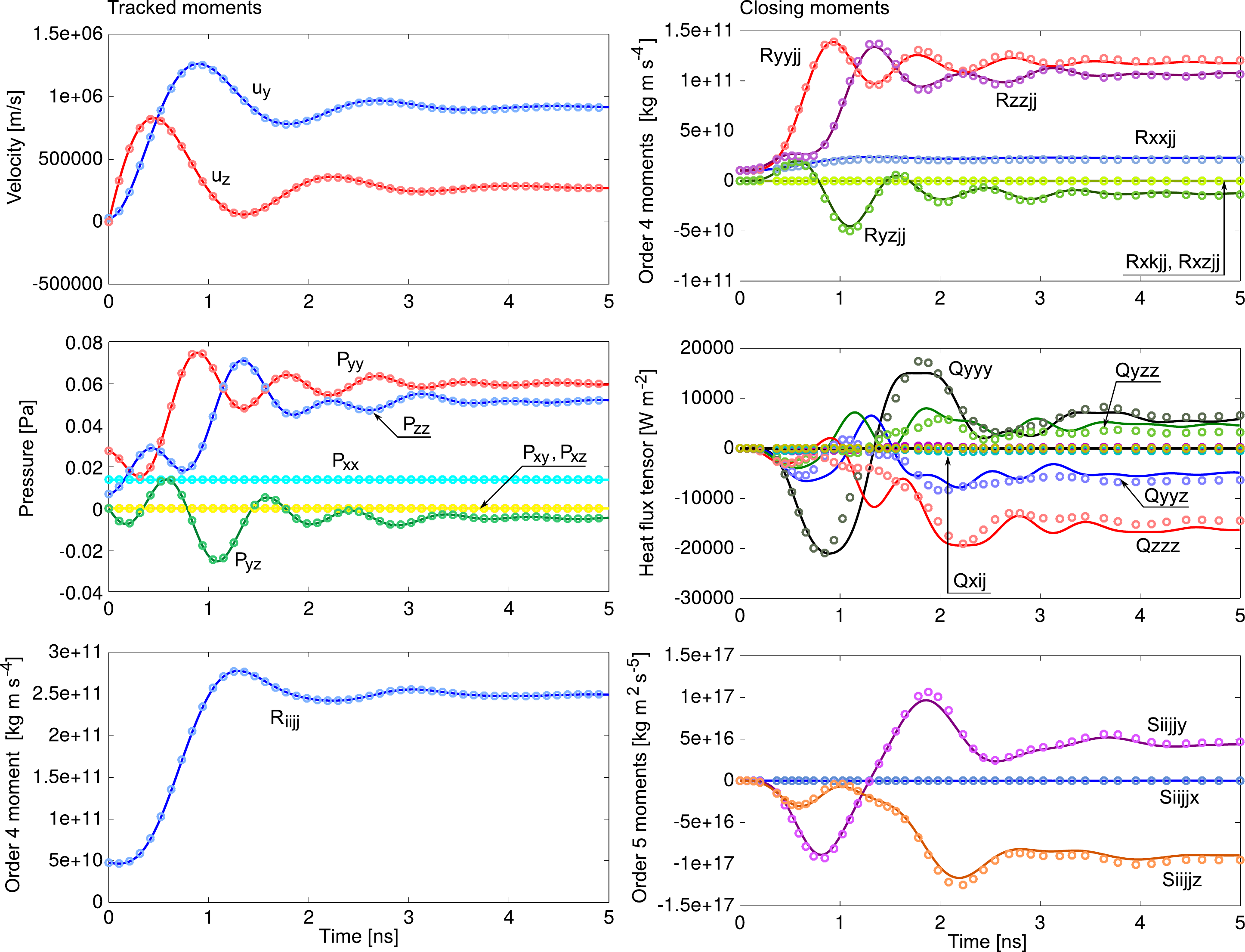}
    \caption{Homogeneous relaxation to a fixed temperature, obtained by a kinetic particle-based method (symbols) and the 14-moment maximum-entropy system with iterative closure (solid lines). 
    Left: some of the tracked moments, reproduced exactly. Right: closing moments. In the labels, $Q_{xij}$ refers to all entries in the heat flux tensor that include the x-velocity component.}
    \label{fig:homogeneous-relaxation-results}
\end{figure*}

We solve this equation with a PIC solver to obtain a reference kinetic solution (more details in Appendix~\ref{sec:appendix-PIC-parameters}).
The 14-moments system of Eq.~(\ref{eq:governing-equations-fluid}) is then solved for the same initial conditions.
Notice that since the problem is space-homogeneous, there are no fluxes in the 14-moment equations, therefore no closure is needed and the 14-moment system is able to retrieve exactly the kinetic solution.
This is shown in Fig.~\ref{fig:homogeneous-relaxation-results}-Left.
It is possible to note how the velocity in the axial direction reaches the classical value given by the mobility from Eq.~(\ref{eq:cross-field-trasport}): $u_z \approx 2.7 \times 10^5 \ \mathrm{m/s}$. 

\noindent The velocity $u_y$ is essentially equal to the value $\mathsf{E}/\mathsf{B} = 10^6 \ \mathrm{m/s}$, slightly reduced by the presence of collisions.
The distribution function from the PIC solver at steady state overlaps with the analytical solution for electrons in uniform fields.\cite{shagayda2012stationary}

Despite not needed by the solution itself, we analyze the suitability of the moment method by computing the closing moments at every timestep, and compare them with the ones obtained from the PIC solver.
Closing moments are obtained by the iterative solution of the entropy-maximisation problem in Eq.~(\ref{eq:minimization-problem-J}).
The result is shown in Fig.~\ref{fig:homogeneous-relaxation-results}-Right and shows a very good agreement for the $R_{ijkk}$ and the $S_{ijjkk}$ moments, and a reasonable reproduction of the $Q_{ijk}$ terms.
This could be due to the fact that $R_{ijkk}$ and $S_{ijjkk}$ contain some contractions over the velocity components, such that some error is removed.
However, it should be remarked that by computing the non-dimensional values as discussed in Eq.~(\ref{eq:nondim-Q-R-S}), the $R_{ijkk}$ and $S_{ijjkk}$ appear to have the same importance, while the $Q_{ijk}$ terms are roughly 10 times smaller.
Therefore, the test case confirms the quality of the 14-moment description.  Finally, the approximated interpolative closure\cite{mcdonald2013affordable} was also tested, giving equally accurate results.


\section{Conclusions}\label{sec:conclusions}

In this study, we investigated the 14-moment closure for the description of electrons in crossed electric and magnetic fields, and developed the required source terms.  As a first step, the 14-moment description with maximum-entropy closure was compared to the steady analytical solution for electrons in uniform fields and in absence of space gradients.
The maximum-entropy closing fluxes show good agreement with the analytical ones, suggesting the applicability of the method to this degree of non-equilibrium.

The 14-moment system was then extended as to include the effect of electric and magnetic terms.
Collision sources were then introduced in a BGK-like form, adapted to describe the impact of hot electrons on a background of stationary neutrals, which is the predominant collision type inside Hall thruster devices.
A second collision model was derived, as to account for the detailed energy conservation during the collision, arising from the large mass disparity between collision partners.

The effect of the electrons VDF on the collision frequency is investigated assuming a simplified cross-section.
The effect of the electrons drift velocity shows to play an important role in the problem, even more significant than the non-Maxwellian shape of the VDF.
An approximated expression for the collision frequency is proposed to represent this contribution, showing good accuracy in the considered conditions.

Finally, the model was applied to the study of 0D relaxation problems, starting from an arbitrary initial anisotropic and drifted velocity distribution function.
Since the considered problem is homogeneous, no space fluxes come into play, and the description results exact.
The value of the maximum-entropy closing fluxes is checked all along the relaxation, and compared with the results from a kinetic simulation, showing the accurate predictions of the 14-moment closure.
  
The results described herein all suggest that the 14-moment closure is capable of reliable descriptions of magnetized electrons.
Future research activities will focus on the implementation of such system in spatially non-uniform conditions in multiple dimensions, as well as on the development of further non-equilibrium source terms for electron-neutral collisions.

The data that supports the findings of this study are available within the article.


\begin{acknowledgements}

The authors wish to thank Prof. Aldo Frezzotti (Politecnico di Milano, Italy), for the insightful discussions on the topic.

\end{acknowledgements}


\appendix


\section{Conserved variables and fluxes}\label{sec:appendix-full-sys14mom}

We provide here the full expression for the 14-moment system.
For simplicity, index notation is employed, where repeated indices imply summation.

\begin{widetext}
\begin{subequations}
\begin{equation}
    \tfrac{\partial}{\partial t} \rho 
    + \tfrac{\partial}{\partial x_i} \left( \rho u_i \right) = S_1 
\end{equation}
\begin{equation}
    \tfrac{\partial}{\partial t} \left( \rho u_i \right)
    + \tfrac{\partial}{\partial x_j} \left( \rho u_i u_j + P_{ij} \right) = S_{2,3,4}
\end{equation}
\begin{equation}
    \tfrac{\partial}{\partial t} \left( \rho u_i u_j + P_{ij} \right)
    + \tfrac{\partial}{\partial x_k} \left( \rho u_i u_j u_k + u_i P_{jk} + u_j P_{ik} + u_k P_{ij} + Q_{ijk} \right) = S_{5-10}
\end{equation}
\begin{multline}
    \tfrac{\partial}{\partial t} \left( \rho u_i u_j u_j + u_i P_{jj} + 2 u_j P_{ij} + Q_{ijj} \right) 
    +\tfrac{\partial}{\partial x_k} \left(\rho u_i u_k u_j u_j  + u_i u_k P_{jj} + 2 u_i u_j P_{jk} + 2 u_j u_k P_{ij} + u_j u_j P_{ik}  \right. \\
    \left.  + u_i Q_{kjj} + u_k Q_{ijj} + 2 u_j Q_{ijk} + R_{ikjj} \right) = S_{11,12,13}
\end{multline}
\begin{multline}
    \tfrac{\partial}{\partial t} \left( \rho u_i u_i u_j u_j + 2 u_i u_i P_{jj} + 4 u_i u_j P_{ij} + 4 u_i Q_{ijj} + R_{iijj} \right)
    + \tfrac{\partial}{\partial x_k} \left( \rho u_k u_i u_i u_j u_j + 2 u_k u_i u_i P_{jj} + 4 u_i u_i u_j P_{jk} \right. \\ 
    \left. + 4 u_i u_j u_k P_{ij} + 2 u_i u_i Q_{jkk} + 4 u_i u_k Q_{ijj} + 4 u_i u_j Q_{ijk} + 4 u_i R_{ikjj} + u_k R_{iijj} + S_{kiijj} \right) = S_{14}
\end{multline}
\end{subequations}
\end{widetext}

For more details, see McDonald \& Torrilhon.\cite{mcdonald2013affordable}


\section{Parameters for the particle simulation}\label{sec:appendix-PIC-parameters}

The particle simulation of Section~\ref{sec:0D-relaxation} were performed using a timestep of $1/(100 \, \nu_c)$ and $10^7$ particles.
The BGK-like collisions are implemented by a simple stochastic method.
For every simulated particle, a collision probability $P_c$ is computed from:
\begin{equation}
    P_c = 1 - \exp(-\nu_c \Delta t)
\end{equation}

\noindent with $\Delta t$ the simulation time step and $\nu_c$ the imposed collision frequency.
The collision happens if a random number $\mathcal{R}$ is smaller than $P_c$.
In such case, the particle velocities are reset to values sampled from the Maxwellian distribution at temperature $T_p$.

The general isotropic collision operator of Section~\ref{sec:isotropic-E-preserving-collis} could be implemented with the same acceptance-rejection procedure, and performing a random rotation of the electron velocity if a collision happens, preserving the initial velocity magnitude.


\nocite{*}

%

%

\end{document}